\documentstyle[twocolumn,aps,prb,epsf,floats]{revtex}

\begin{document}

{\bf \ Schmalian and Wolynes Reply:} The numerical results presented by
Grousson {\em et al.} \cite{Gr} confirm that the model\ we studied in Ref. 
\cite{SW} exhibits self-generated glassiness, which was the major claim of
Ref.\cite{SW}. The decrease of the freezing temperature with increasing
frustration parameter, $Q$, (as inferred from Fig. 1 of Ref.\cite{Gr}) also
agrees with our prediction. Furthermore, Grousson {\em et al. } show that
the temperature dependence of the slow relaxation time can indeed be well
described using the Vogel-Fulcher law which we proposed based on an entropic
droplet argument. Finally, analyzing the $Q$-dependence of the fragility
parameter $D$, Grousson {\em et al. }find that it behaves similar to $D\sim
Q^{1/2}$ and argue, assuming $D^{-1}\sim \left. \frac{dS_{c}}{dT}\right|
_{T_{K}}$ with configurational entropy $S_{c}$ and Kauzmann temperature $%
T_{K}$, that this disagrees with our theory which gives $\left. \frac{dS_{c}%
}{dT}\right| _{T_{K}}\rightarrow 0$ as $Q\rightarrow 0$.

\ We did not study the $Q$-dependence of $D$ in our paper. Using simple
estimates for $D\left( Q\right) $ we find that there actually is no
scientific disagreement between the results of Ref.\cite{Gr} and our theory.
Indeed, we showed in our Letter that $\left. \sim \frac{dS_{c}}{dT}\right|
_{T_{K}}$ vanishes as $Q$ goes to zero. For glass forming molecular liquids $%
\left. \frac{dS_{c}}{dT}\right| _{T_{K}}$ and the parameter $D^{-1}$ are
proportional, both empirically and theoretically\cite{XW}. Thus it is
natural to conclude that $D$ itself must grow for decreasing $Q$. However,
for the stripe glass system discussed in our paper we also pointed out (see
below Eq. 10 of our Letter) that the coefficient $D$ in question is not only
determined by the configurational entropy but also by the bare surface
tension of entropic droplets, $\sigma _{0}$, via 
\begin{equation}
D\sim \frac{\sigma _{0}^{2}}{\left. \frac{dS_{c}}{dT}\right| _{T_{K}}}.
\label{1}
\end{equation}
$\sigma _{0}$ itself depends strongly on the frustration parameter and
overshadows the $Q$-dependence of $\left. \frac{dS_{c}}{dT}\right| _{T_{K}}$
in Eq.\ref{1}. Thus, we find that $D\left( Q\rightarrow 0\right) \rightarrow
0$, in agreement with the results of Ref.\cite{Gr}.

We now summarize our calculations for the $Q$-dependence of $\left. \frac{%
dS_{c}}{dT}\right| _{T_{K}}$ and $\sigma _{0}$. \ Using an analytical theory 
\cite{SWW}, which agrees quantitatively very well with the numerical
solution of the self consistent screening approximation \cite{SW}, gives $%
S_{c}\left( T_{A}\right) \sim Q^{3/4}$. Since $S_{c}\left( T_{K}\right) =0$
we estimate with good accuracy $\left. \frac{dS_{c}}{dT}\right|
_{T_{K}}\simeq S_{c}\left( T_{A}\right) /\left( T_{A}-T_{K}\right) $. $%
T_{A}-T_{K}$ weakly depends on $Q$ for larger $Q$ and increases due to
additional logarithmic corrections for small $Q<0.04$ . It follows that $%
\left. \frac{dS_{c}}{dT}\right| _{T_{K}}\sim Q^{\tau }$ with $\tau =\frac{3}{%
4}$ for $\ $larger $Q$ and a behavior which is closer to $\tau \simeq 1$ for
smaller $Q$ (logarithmic corrections actually spoil a pure power-law, an
effect which is however very hard to identify from a plot of $\left. \frac{%
dS_{c}}{dT}\right| _{T_{K}}$).

Next we estimate $\sigma _{0}\left( Q\right) $ for small $Q$. In this limit
there is, in addition to the usual gradient term (studied in Ref.\cite{SW}
to analyze $\sigma _{0}$ for larger $Q$) a contribution to the surface
energy which results from the Coulomb term of the Hamiltonian, $E_{{\rm %
surface}}^{C}\sim \varphi _{0}^{2}l_{m}^{-1}R^{2}\ $. Here, $\varphi _{0}$
is a typical local amplitude of the charge field of the model, $R$ is the
droplet radius and $l_{m}\sim Q^{-1/4}$ is the modulation length of the
state \cite{Nus}. The $Q$ dependence of $\varphi _{0}$ can be obtained from
a variational argument: we make the ansatz $\varphi \left( r\right) =\varphi
_{0}\cos \left( rQ^{1/4}\right) $ for a modulated structure, insert it into
the Hamiltonian and minimize with respect to $\varphi _{0}$. It follows $%
\varphi _{0}^{2}\sim Q^{1/2}$, a result which we recently used \ in Ref.\cite
{SWW}. This gives a bare surface tension which behaves as $\sigma _{0}\sim
\varphi _{0}^{2}l_{m}^{-1}\sim Q^{3/4}$ as $Q\rightarrow 0$. It then follows
from Eq.\ref{1} that 
\begin{equation}
D\left( Q\right) \sim Q^{\mu }
\end{equation}
with $\mu =\frac{3}{2}-\tau \simeq \frac{1}{2}$ if one uses the effective
exponent $\tau \simeq 1$ for the slope of the configurational entropy at
smaller $Q$. We conclude that $D$ grows in a fashion similar to a square
root for small $Q$ (a pure power-law is spoiled by logarithmic corrections
). Certainly, a more sophisticated replica-instanton calculation is\ called
for to solidify these estimates. Nevertheless, the presented calculation
seems to agree rather well with the simulations presented by Grousson {\em %
et al.}\cite{Gr}.

The claim by Grousson {\em et al.}\cite{Gr} that our theory disagrees with
their numerical simulations is based on the assumption that the surface
tension $\sigma _{0}$ is independent of the frustration parameter $Q$.
However, we showed in this Reply that $\sigma _{0}$ varies strongly with $Q$
such that $D^{-1}$ and $\left. \frac{dS_{c}}{dT}\right| _{T_{K}}$ become
independent measures of the fragility. We believe that this answers the
questions raised in Ref.\cite{Gr}. \ What is great about these stripe
glasses is they present the possibility of experimentally  decoupling the
fragility as measured from thermodynamics and as measured from kinetics,
which has proved impossible for molecular liquids. We are grateful to
Grousson {\em et al.} that they brought up this issue.

This work has been supported by NSF (P. G. W.), Grant No. ChE-9530680 and
Ames Laboratory, operated for the U. S. Department of Energy by Iowa State
University under Contract No. W-7405-Eng-82 (J. S.). We acknowledge helpful
discussions with H. Westfahl Jr.

\bigskip \noindent J\"{o}rg Schmalian,

{\footnotesize Department of Physics and Astronomy and }

{\footnotesize Ames Laboratory, Iowa State University, Ames, IA 50011}

\noindent Peter G. Wolynes,

{\footnotesize Department of Chemistry and Biochemistry, }

{\footnotesize University of California at San Diego, La Jolla, CA 92093}

\end{document}